\documentclass[pre,a4paper,final]{revtex4}
\usepackage{graphicx}
\usepackage{amsmath}
\usepackage{psfrag}
\usepackage{latexsym}

\newcommand{\bq}{\begin{equation}}
\newcommand{\eq}{\end{equation}}
\newcommand{\ba}{\begin{eqnarray}}
\newcommand{\ea}{\end{eqnarray}}

\begin{document}
\title{Balls-in-boxes condensation on networks}
\author{L. Bogacz$^{1,3}$}{\thanks{bogacz@th.if.uj.edu.pl}
\author{Z. Burda$^{1,2}$}{\thanks{burda@th.if.uj.edu.pl}
\author{W. Janke$^3$}{\thanks{Wolfhard.Janke@itp.uni-leipzig.de}
\author{B. Waclaw$^{1,3}$}\thanks{bwaclaw@th.if.uj.edu.pl}
\address{$^1$Marian Smoluchowski Institute of Physics,
Jagellonian University, Reymonta 4, 30-059 Krak\'ow, Poland, \\
$^2$Mark Kac Complex Systems Research Centre, Jagellonian University, Krak\'ow,
Poland \\
$^3$Institut f\"ur Theoretische Physik and Centre for
Theoretical Sciences (NTZ), Universit\"at Leipzig,
Augustusplatz 10/11, 04109 Leipzig, Germany}

\begin{abstract}
We discuss two different regimes of condensate formation
in zero-range processes on networks:
on a $q$-regular network, where the condensate
is formed as a result of a spontaneous symmetry breaking,
and on an irregular network, where the symmetry
of the partition function is explicitly broken. 
In the latter case we consider a  minimal irregularity 
of the $q$-regular network introduced by a single $Q$-node with degree $Q>q$. 
The statics and dynamics of the condensation depends on
the parameter $\alpha = \ln Q/q$, which
controls the exponential fall-off of the distribution of particles 
on regular nodes and the
typical time scale for melting of the condensate
on the $Q$-node which increases exponentially with the system 
size $N$.
This behavior is different than that on a $q$-regular network 
where $\alpha=0$ and where the condensation results from the
spontaneous symmetry breaking of the partition
function, which is invariant under a permutation of
particle occupation numbers on the $q$-nodes of the network.
In this case the typical time scale for condensate melting is known
to increase typically as a power of the system size. 

\bigskip

\end{abstract}
\maketitle

The formulation of the principles of non-equilibrium
statistical mechanics is a challenge for theoretical physics.
Non-equilibrium effects play an important role in many phenomena 
but we do not have a consistent theory which would describe
them. An exception may be systems close to equilibrium for which we
can gain some insight into their dynamics
using ideas of the linear response and of the 
fluctuation-dissipation theorem. The far-from-equilibrium dynamics is 
an uncharted territory. It is therefore useful to look 
for solvable models belonging to the latter class, which could
teach us about what happens in this case.

In this note we shall discuss zero-range processes
on networks which belong to this class. On the one hand,
they reveal an interesting, non-trivial and very rich behavior 
including a condensation, far-from-equilibrium dynamics and
non-linear effects as for example the formation or melting of 
the condensate. On the other hand, due to a relation 
to the balls-in-boxes model \cite{bbj,grav}, which is analytically solvable, 
also these non-trivial effects are analytically treatable.

The zero-range process is a stochastic process which describes
a gas of identical particles hoping between neighboring sites 
of a lattice or network on which the particles reside 
(for reviews, see \cite{godreviews, evans, evans2}).
The transition rate for particles to hop from one site 
to an adjacent site depends only
on the state of the node from which the particle hops
and is independent either of the destination node or any
other node. Since the hop rate requires only ultra-local information,  
the corresponding process is called zero-range process. 

When the density of particles exceeds a certain critical value,
the system undergoes a condensation \cite{bbj}
in which a condensate is formed
on a single node of the network which attracts a large
number of particles. The condensation takes place in real space
and not in momentum space as for the Bose-Einstein 
condensation. This type of condensation is called balls-in-boxes 
condensation, in short B-in-B or backgammon condensation because
it was first discovered in the balls-in-boxes (backgammon) model
\cite{bbj,grav}.

The B-in-B condensation is observed in many systems. For example
in statistical models of random trees (also called branched
polymers) one observes a phase transition between a phase 
of generic elongated trees to a phase where a typical
tree looks rather like a bush with a singular node which has
a finite fraction of all branches \cite{bbj,trees}. The statistics
of the degree distribution in the tree model can be mapped
onto the B-in-B model. The emergence
of a singular node on a tree corresponds
to the emergence of a condensate of many balls in one box in
the B-in-B model.
A similar geometrical phase transition is observed in models 
of quantum manifolds discussed in the 
context of quantum gravity \cite{grav1,grav2}. 
At the phase transition the quantum manifolds collapse to 
a very singular geometry 
whose volume is almost fully concentrated in the closest neighborhood 
of a single point of the manifold. More precisely, the ratio $v/V$
of the volume $v$ of the neighborhood 
within a radius of order of an ultraviolet cut-off around this point
to the total volume $V$ of the whole manifold is finite 
in the limit $V \rightarrow \infty$ even if the
radius (ultraviolet cut-off) is kept constant. 
Again this phenomenon can be viewed as a particular realization 
of the B-in-B condensation \cite{grav1}.

The B-in-B condensation explains
many other phenomena as for instance wealth condensation \cite{wealth},
emergence of singular nodes in complex networks \cite{cn1,bbw}, 
emergence of the Hagedorn fireball in hadron physics \cite{haged1,haged2} and
some transitions observed in shaken granular gases \cite{shaken,shaken2} or 
the phase transition in zero-range processes \footnote{In one dimension
a zero-range process can be mapped onto an asymmetric 
exclusion process \cite{evans}.}. Actually, the mathematics of 
the condensation is also almost identical to that of the Berlin-Kac 
phase transition in the spherical model \cite{kac}. In other words
the B-in-B mechanism is quite generic and common.

The state of the zero-range process is characterized by the distribution 
of the numbers of particles at all $N$ nodes of the network:
$\{m_i\} = \{m_1,\dots, m_N\}$ where $m_i$ denotes the
occupation number of the $i$-th node. 
The total number of particles $M=m_1+\ldots+m_N$ is constant during
the process. Particles hop from non-empty sites 
with a rate $u(m_i)$ depending only on the site
occupation number $m_i$. 
The outgoing current of particles from a site $i$
is distributed equally among all $q_i$ links emerging 
from the node, so the effective
hop rate per link is $u(m_i)/q_i$, where $q_i$ is called node degree. The function $u(m)$ 
is identical for all nodes, but the factor $1/q_i$ is not
since it explicitly depends on the node degree.
In this note we assume that the network topology 
is fixed and so is the degree sequence $\{q_i\}$.
The most fundamental question is whether the process
has a steady state and if so, whether it is unique. 
The answer to this question is affirmative if 
the network is connected. In this case the process has 
a unique steady state 
which depends only on the node degrees $\{q_i\}$ and 
the numbers of nodes $N$ and particles $M$. 
This steady state corresponds to the only equilibrium state
which is sooner or later reached by the process. In equilibrium, the 
partition function can be calculated analytically \cite{evans}: 
\bq
Z = \sum_{m_1=0}^M \cdots \sum_{m_N=0}^M 
\delta_{m_1+\dots+m_N,M}
\prod_{i=1}^N p(m_i) q_i^{m_i} ,
\label{part}
\eq
where
\bq
p(m)= \prod_{k=1}^m \frac{1}{u(k)}, \;\; p(0)=1 .
\label{pbyu}
\eq
The weight $p(m)$ is identical for every node, but 
the total node weights $p_i(m) \equiv p(m) q_i^m$ have an 
additional contribution $q_i^m$ explicitly depending 
on the node degree. Because of the presence of the
(Kronecker) delta function under the sum in Eq. (\ref{part}), 
the partition function does not entirely factorize 
into a product of independent weights for individual nodes. 
The constraint on the 
total number of particles plays an important role as we shall
see below, because the occupation numbers of individual nodes 
are not independent of each other. 

The statics of the zero-range process
is equivalent to a B-in-B model 
with the partition function (\ref{part}) 
describing a system of $M$ identical balls distributed 
in $N$ boxes, each having a weight function $p_i(m)$.
The probability that in equilibrium the system is 
in a state $\{m_i\}$ reads: 
\bq
P(m_1,\dots,m_N) = \frac{1}{Z} \prod_{i=1}^N p(m_i) q_i^{m_i} 
= \frac{1}{Z} \prod_{i=1}^N p_i(m_i),
\label{Pdistr}
\eq
where as before $m_1+\cdots+m_N=M$.
It is interesting to notice that this probability is
invariant with respect to the following change of the 
weights:
\bq
p(m) \rightarrow C e^{\mu m} p(m).
\label{inv}
\eq
Indeed under the change (\ref{inv}) the partition function (\ref{part}) 
changes as $Z \rightarrow C^N e^{\mu M} Z$. The multiplicative
factor which appears in front of the partition function
cancels out in the probability (\ref{Pdistr}) and thus the 
statistical averages do not depend on the parameters $C$ and $\mu$. 
This invariance is an important property of the model. 
The parameter $C$ has the meaning of a normalization and can
be used for example to normalize the weights to a probability.
One should notice that the hoping rate $u(m)=p(m-1)/p(m)$ is
not affected by $C$. The parameter $\mu$ or more specifically
$e^{\mu}$ rescales the hoping rate $u(m) \rightarrow e^{-\mu} u(m)$
or equivalently stretches the time scale $t\to t e^\mu$.

The probability distribution (\ref{Pdistr})
encodes the full information
about the static properties of the system. For
example one can calculate statistical averages of any
observable $X$:
\bq
\langle X \rangle =  \sum_{\{m_i\}} P(m_1, \ldots, m_N ) 
X(m_1,\ldots,m_N)
\eq
or correlations $\langle XY \rangle -\langle X \rangle \langle Y \rangle$
etc. A particularly interesting observable is the number of incidents 
that the $i$-th node is occupied by $m$ particles: 
\bq
\pi_i(m) = \left\langle \delta_{m_i,m} \right\rangle = p_i(m) \frac{Z_i(m)}{Z}
\label{pii}
\eq
where, using the B-in-B analogy,
$p_i(m)$ is the weight of the $i$-th box, $Z$ is the partition
function for the total system of $M$ balls in $N$ boxes
and $Z_i(m)$ is the partition function for $M\!-\!m$ balls in 
the $N\!-\!1$ remaining boxes. 
In a similar manner one can count multi-node distributions
and average them over all configurations.
For example averaging the two-node incident function
$\delta_{m_i, m} \delta_{m_j,n}$ over all configurations
gives the probability $\pi_{ij}(m,n)$ that in equilibrium
there are simultaneously $m$ particles at node $i$ and 
$n$ particles at node $j$.

A zero-range process is said to be in the condensed phase
if a finite fraction of all particles tends to occupy one node
or, if one phrases it in terms of the underlying B-in-B model, 
if a finite fraction of all balls is concentrated in one 
box. The effect depends on the density of 
particles per node (or balls per box) 
$\rho = M/N$. If one keeps $\rho$ constant and 
takes the limit $N\rightarrow \infty$,
one sees that above a certain critical density $\rho_{c}$ 
one box contains on the average a large number of balls $(\rho-\rho_c) N$ which
grows with $N$, while any other box has only $\rho_{c}$ balls.

The most thoroughly studied and probably most surprising example
of the B-in-B condensation is the condensation which
takes place in a system of identical boxes, that is
for which all the weight functions in Eq. (\ref{Pdistr}) are
identical: $p_1(m) =\dots =  p_N(m)$. The corresponding zero-range process 
is realized on a $q$-regular graph which has
identical degrees of all nodes. In this case the partition
function (\ref{part}) is symmetric with respect to the
permutation of the box-occupation numbers $\{m_i\}$. The condensation
appears there as a result of a spontaneous symmetry breaking
which selects one out of identical boxes for the location
of the condensate. The criterion for the appearance of
the condensation is controlled by the asymptotic value 
of the hop rate $u(m)$ for $m\rightarrow \infty$, 
\bq
\frac{p(m-1)}{p(m)} = u(m) \rightarrow u_\infty .
\label{q_u}
\eq
For $u_\infty \! = \! \infty$ 
the corresponding critical density 
is infinite \footnote{Or if $p(m)=0$ for all $m$ larger
than a certain $m_0$.} 
and there is no condensation in the model.
The system is always in a fluid phase. 
This can be intuitively understood because if the hop rate $u(m)\to\infty$,
it amounts to an effective repulsion between particles which in effect
avoid to occupy the same site. On the contrary, for 
$u_\infty \!= \! 0$ there is no price whatsoever for particles to pay
for a numerous occupation of the same site, so in effect
the particles tend to condense. The critical density is zero in this
case and therefore the system can be then only in the condensed phase. 
One can say that due to an effective attraction, particles tend 
to keep as close as possible to each other. The most interesting 
case is when $u_\infty$ is finite: 
$0\!<\! u_\infty \!<\!\infty$. 
Actually it is sufficient to consider only the case $u_\infty\!=\!1$ 
since within this model all other values of $u_\infty$ are 
equivalent to one 
as follows from the invariance with 
respect to the transformation (\ref{inv}). 
In the remaining part of the paper we shall therefore stick 
to the choice $u_\infty\!=\!1$. 
The quantitative behavior depends on the exact form of $u(m)$  
but the critical properties, such as critical
exponents, depend only on how $u(m)$ approaches unity when $m$ goes to
infinity. When it behaves as $u(m) = 1 + \beta/m + \ldots$ 
for large $m$, then $p(m)$ behaves asymptotically as
$p(m)\sim m^{-\beta}$. In particular, one can choose $p(m)$ to be
\bq
p(m) = \frac{(\beta-1)\Gamma(\beta)m!}{\Gamma(\beta+m+1)} 
\sim \frac{(\beta-1)\Gamma(\beta)}{m^{\beta}}. \label{pmgamma}
\eq
When the density $\rho=M/N$ exceeds the critical value $\rho_{c}$, 
a condensate with $N\Delta\rho$ particles is formed \cite{evans,god}, 
where $\Delta\rho=\rho-\rho_c$. The critical density is given by the formula
\bq
\rho_c = \frac{\sum_{m=0}^{\infty} m p(m)}{\sum_{m=0}^{\infty} p(m)}
\eq
and can be concisely expressed in terms of the generating function 
$K(\mu) = \sum_{m} p(m) e^{-\mu m}$
as $\rho_c = -K'(0)/K(0)$. In particular, for 
$p(m)$ as in Eq. (\ref{pmgamma}), the critical density 
is $\rho_c=1/(\beta-2)$. An interesting choice of the weights $p(m)$
is when one demands that every box has at least one particle, 
and $p(m)=m^{-\beta}$. In this case the generating function has 
the following integral representation:
\bq
K(\mu) = \frac{1}{\Gamma(\beta)}
\int_0^{\infty} {\rm d} t \ \frac{t^{\beta-1}}{e^{\mu+t}-1},
\eq
which uncovers mathematical similarities between the B-in-B
and the Bose-Einstein condensation. The statics of the model 
can be solved analytically \cite{bbj,grav}, with critical
properties depending on $\beta$.

The dynamics of the condensation has also been studied analytically 
\cite{evans,god,cc4}. Two questions can be posed: what is the typical 
time scale for building the condensate from a homogeneous distribution 
of particles and what is its average life time. Here we shall focus 
on the latter. Once the condensate is formed, it moves across the system. 
It spends a long time at one particular site
but sometimes melts and is rebuilt at another site. 
A typical time scale for melting the condensate 
has been derived using mean-field arguments \cite{god}. 
In the mean-field approach one monitors only
a single node of the network and derives 
effective equations balancing 
the in-flow and out-flow of particles for this node.
One does not care about what happens in the remaining part 
of the system which in this approximation
is treated as a homogeneous reservoir of particles, 
where fluctuations are much faster than
the dynamics of the condensate. 
The monitored node is characterized 
by the distribution of the number of particles $\pi_i(m)$ which,
in an adiabatic approximation, is assumed to be that of the steady state.
For a homogeneous system, $\pi_i(m)$ is identical
for all nodes, so it is equal to the average over all nodes:
$\pi_i(m)=\overline{\pi_i}(m) \equiv \pi(m)$, and it does not 
matter which node is monitored.
The full information is encoded in $\pi(m)$. 
The occupation number of the monitored node may change in
one step by one unit or stay constant. This sequence of changes 
can be viewed as a random walk in the effective one-dimensional
potential $V(m) = -\ln \pi(m)$.
The waiting time for a condensate to melt can be thus
viewed as the time needed for a particle 
to randomly walk from $m_*$, which corresponds to the
value of the condensate $m_*\equiv N\Delta \rho$, to some $m_0 \ll m_*$.
This time is related to going through
the maximum of the potential, whose position 
corresponds to the position of the dip of the function 
$\ln \pi(m)$. This position is very close 
to $m_*/2$ because the excess of particles is shared mainly 
by two nodes \cite{god}. In the condensed phase the shape
of the function $\pi(m)$ can be approximated in the range
of $m$ from zero to the dip location (see Fig. \ref{fig1}) 
by $\pi(m) \approx p(m)$, where we assume that $\sum_m p(m)=1$ 
as follows from Eq. (\ref{pmgamma}). So  
the value of the function $\pi(m)$ at the dip
is roughly equal to $p(m_*/2)$ which gives
the corresponding maximum of the effective potential 
$V_{*} = -\ln p(m_*/2)$. Thus using the Arrhenius law
one can expect that the time needed for a random walk to
go over this maximum is of order $\tau \sim e^{V_*} = 1/p(m_*/2)$. 
For $p(m)$ asymptotically behaving as $\sim m^{-\beta}$
this yields $\tau \sim N^\beta$, i.e., a power law in the system size $N$,
with a coefficient 
proportional to $\Delta \rho^\beta$. This crude argument
gives already a good estimate. It can be polished
if one implements all details of the zero-range dynamics
into the mean-field analysis and works out the consequences
of the detailed balance condition for the transition rates 
for a particle to hop into or from the monitored node or to stay at it
\cite{godreviews,god}.
One obtains an expression for a monitored node $i$ for the
waiting time $\tau^i_{k\rightarrow m}$ which tells us how long
it takes to fall from $k$ to $m<k$ particles, where only 
the first-passage time is taken into account:
\bq
\tau^i_{k\rightarrow m} = 
\sum_{r=k+1}^{m} 
\frac{1}{u(r)\pi_i(r)} \sum_{l=r}^{M} \pi_i(l)  .
\label{Tmngen}
\eq
For a regular graph the distribution $\pi_i(m)$ is identical
for each node so the index $i$ can be skipped. The typical melting time,
defined to be $\tau = \tau_{m_{*}\rightarrow 0}$,
calculated from this equation grows as before as
$\tau \sim N^\beta$, but with a slightly
modified coefficient which is now proportional
to $\Delta \rho^{\beta+1}$. It is only a small correction
to the previously derived result since the dependence on $N$ is the 
same. 

\begin{figure}
\psfrag{yy}{$\pi(m)$} \psfrag{xx}{$\frac{m}{M}$}
\includegraphics*[width=10cm]{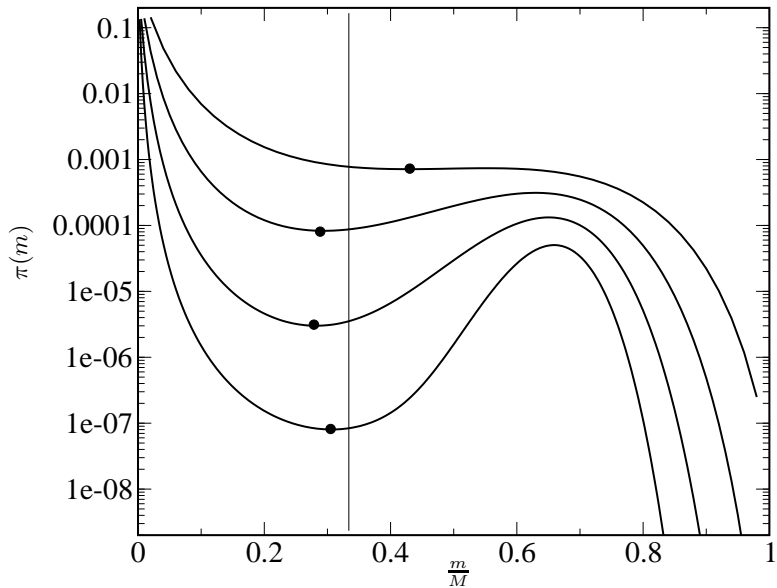}
\caption{Plots of the distribution $\pi(m)$ in the condensed phase, 
for $\beta=5$ and $\rho=1$. The critical density is $\rho_c=1/3$.
From top to bottom: $N=M=50,100,200,400$. The position of the dip 
is marked by a filled circle on each curve. The vertical line denotes 
its asymptotic position as $N\to\infty$: $m_*/2M = 1/3$.}
\label{fig1}
\end{figure}

In contrast to homogeneous systems, much less is known about
the zero-range process on inhomogeneous networks,
where the symmetry of the partition function (\ref{part}) resulting
from identical node degrees is explicitly broken. 
Some attempts have been made for scale-free networks
\cite{jdn} to explore the phase diagram and the dynamics 
of condensate formation, but not the life time of the condensate.

We will now take advantage of Eq. (\ref{Tmngen})
to estimate the typical time scale for the dynamics of
condensation for inhomogeneous systems, 
assuming that (\ref{Tmngen}) applies within the scope of the
mean-field approximation also when $\pi_i(m)$ vary from node to node.  
We will study the effect of inhomogeneity by introducing to 
a $q$-regular graph  a single node with degree $Q>q$. This type 
of irregularity with a single node being different from the others is well
suited to study the effect of symmetry breaking in zero-range
processes, which generates a condensate on a single node. 
The parameter $\alpha = \ln Q/q$ plays the role of an
external field which breaks this symmetry.
Let us first determine the static properties of such a system. 
Because the node weight for the $Q$ node, $p_Q(m) = Q^m p(m)$, differs
from the weights for regular nodes, $p_q(m) = q^m p(m)$, by an 
exponential factor $(Q/q)^m$, it clear that
this node has a tendency to attract more particles than the others. 
We hence expect that $\pi_Q(m)$ increases fast with $m$, while 
$\pi_q(m)$ decreases.
The exact form of these distributions depends on the particular form of
the weight function $p(m)$ in Eq. (\ref{part}) 
but this does not significantly change the generic behavior. 
The exponential effect coming from the factor $Q/q > 1$ is dominant. 
In order to simplify the calculations it is therefore convenient to assume
that the transition rates do not depend on $m$: $u(m)=1$ 
or equivalently $p(m)=1$. 
Assuming this, one can show \cite{ournew}, that the distribution
$\pi_Q(m)$ develops a maximum for $m$ close to the upper limit $M$:
\bq
\pi_Q(m) \propto \left(\frac{Q}{q}\right)^m \binom{M+N-m-2}{M-m}, \label{piQ}
\eq
where we have skipped an overall normalization.
As an example, we show in Fig. \ref{fig2} the distribution
$\pi_Q(m)$ of particles at the $Q$-node. Above the critical 
density $\rho_c=1/(\frac{Q}{q}-1)$, the distribution
has a maximum for a number of particles $m_*$ which 
linearly grows with the total system size, $m_* \sim N$, so
particles condense at the $Q$-node. In the same figure
we also show the distribution of particles on a regular
node $\pi_q(m)$, where it falls off exponentially:
\bq
	\pi_q(m) \propto \left(\frac{q}{Q}\right)^m.
\eq
We can now calculate the typical time scales 
$\tau_{Q} = \tau^Q_{m_*\rightarrow 0}$ for the condensate to disappear
for the first time from the $Q$-node or 
$\tau_{q} = \tau^q_{m_*\rightarrow 0}$ for the same quantity 
at the regular $q$-node, applying Eq. (\ref{Tmngen}).
The general expression for $\tau_{Q}$ is quite 
complicated \cite{ournew}. In Fig.~\ref{fig3} we plot 
this quantity for $N=20$ and various $M$.
In the thermodynamic limit for fixed density 
$\rho$ it simplifies to
\bq
	\tau_{Q} \sim \exp 
\left\{N\left[ \rho\log\rho+\rho\alpha -(1+\rho)\log(1+\rho)\right]\right\}.
\eq
Furthermore, for $\rho\gg 1$ we get $\tau_{Q} \propto  e^{\alpha \rho N}$
which means that here the characteristic ``melting time'' grows 
exponentially with the system size, while $\tau_{q}$ 
is found to grow only linearly.
Actually the time needed for the condensate to evaporate from
the $Q$-node can be estimated also using the Arrhenius law, as before,
if one applies it to the effective potential
$V(m) = -\ln \pi_Q(m)$.
As follows from Eq.~(\ref{piQ}), this function has its maximum at $m=0$ and the 
value of this maximum, when one normalizes (\ref{piQ}), can be estimated from
the Arrhenius law: 
$\tau_* \approx e^{V(0)} = 1/\pi_Q(0) \approx e^{\alpha \Delta \rho N}$. 
So in contrast to the homogeneous case the time
grows now exponentially with the system size.

\begin{figure}
\psfrag{xx}{$\frac{m}{M}$} \psfrag{yy}{$\pi(m)$}
\includegraphics*[width=10cm]{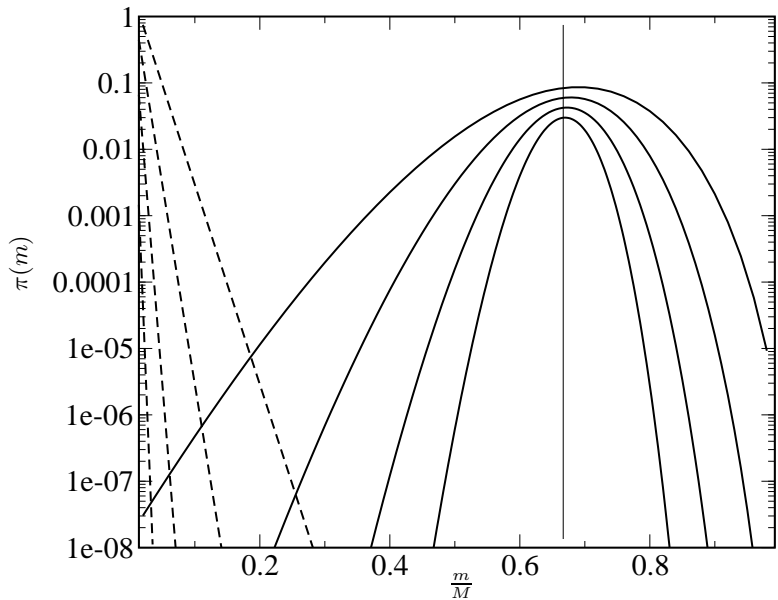}
\caption{Plots of $\pi_Q(m)$ (solid lines) and $\pi_q(m)$ (dashed lines) for the almost $q$-regular graph with one node $Q>q$. Here $q=4,Q=16,\rho=1$ and the critical density $\rho_c=1/3$. The curves from top to bottom show: $N=M=50,100,200,400$. The vertical line marks the asymptotic position of $m_*/M=2/3$.}
\label{fig2}
\end{figure}

\begin{figure}
\psfrag{xx}{$M$} \psfrag{yy}{$\tau_{Q}$}
\includegraphics*[width=10cm]{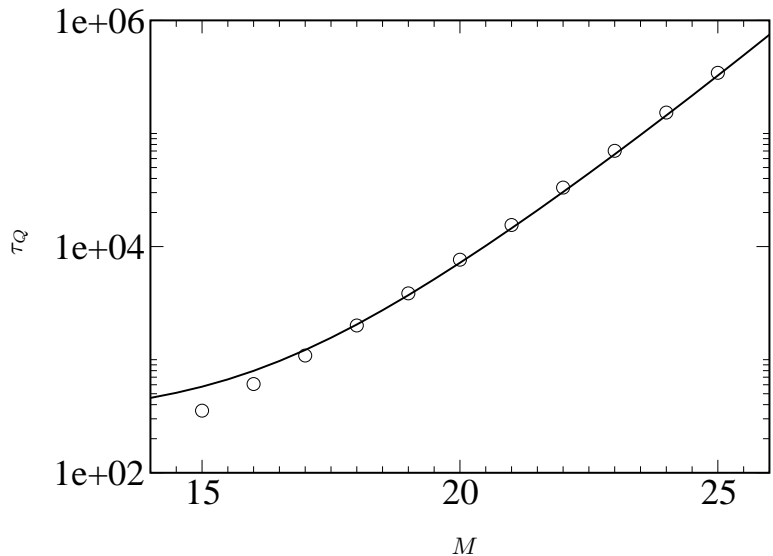}
\caption{The typical time scale $\tau_{Q}$ related to the average life time of the condensate, for $N=20,q=4,Q=16$ and different $M$. 
Circles denote experimental points.}
\label{fig3}
\end{figure}

Let us summarize differences between 
the condensation observed in zero-range processes on a
$q$-regular network and an irregular network.
In the first case the partition function is symmetric
with respect to permutations of the occupation numbers $\{m_1,m_2,\ldots,m_N\}$ 
so that any distribution of particles 
is as probable as any other. This
symmetry is spontaneously broken at the critical point
where a single node containing the condensate 
becomes different from the others. The symmetry
is reduced to the group of permutations
of the remaining $N\!-\!1$ nodes. On an irregular network, on the other hand,
the symmetry is explicitly broken. Because of the nature
of the symmetry breaking, which produces a condensate
on a single node, one can expect that already a model
with an irregularity on a single node is sufficient
to capture the main characteristics of this transition. 
We have studied such a minimal irregularity 
coming from a single node with degree $Q>q$. The parameter
$\alpha = \ln Q/q$ plays the role of an external field.
The situation is very similar as for standard phase
transitions. For $\alpha>0$, one observes a characteristic
exponential suppression $\pi_q(m) \sim e^{-\alpha m}$ 
of the particle distributions on $q$-regular nodes which
can be compared to the exponential fall-off of
the two-point correlation function in standard field theoretical
models, while for the absence of the external field 
one observes long range fluctuations:
$\pi_q(m) \sim p(m)$ which for the interesting
class of weights $p(m)$ is of the power-law type:
$\pi_q(m) \sim m^{-\beta}$. 
This change of behavior has also immediate implications
for the dynamics: a typical time scale for the condensate
melting for $\alpha>0$ grows exponentially with the system
size and for $\alpha=0$ sub-exponentially, typically as a power.

A next step is to consider zero-range processes on complex networks
with an arbitrary degree distribution. One can in particular
address the question of self-averaging that is whether a zero-range
 process on a single typical network chosen at random 
from the given ensemble of networks with the given degree 
distribution behaves in the same way as the corresponding
process averaged over many networks from this ensemble. 
This would be the first step towards the investigations 
of dynamics on networks coupled to network topology.

\section*{Acknowledgments}
This work was supported in part by the EC-RTN Network ``ENRAGE'' under 
grant No. MRTN-CT-2004-005616, an Institute Partnership grant 
of the Alexander von Humboldt Foundation,
an EC Marie Curie Host Development Fellowship under 
Grant No. IHP-HPMD-CT-2001-00108 (L. B. and W. J.) and an EC ToK 
Grant MTKD-CT-2004-517186 ``COCOS'' (Z. B.).
B. W. acknowledges support by a Fellowship of the German Academic 
Exchange Service (DAAD).


\begin{thebibliography}{99}
\bibitem{bbj} P. Bialas, Z. Burda and D. A. Johnston, 
Nucl. Phys. B {\bf 493}, 505 (1997). 
\bibitem{grav} P. Bialas, Z. Burda and D. A. Johnston, 
Nucl. Phys. B {\bf 542}, 413 (1999).
\bibitem{godreviews} C. Godr\'{e}che, cond-mat/0603249; cond-mat/0604276.
\bibitem{evans} M. R. Evans and T. Hanney, J. Phys. A: Math. Gen. {\bf 38},
R195 (2005).
\bibitem{evans2} M. R. Evans, S. N. Majumdar and R. K. P. Zia, 
J. Stat. Phys. {\bf 123}, 357 (2006).
\bibitem{trees} P. Bialas and Z. Burda, Phys. Lett. B {\bf 384}, 75 (1996).
\bibitem{grav1} P. Bialas, Z. Burda, B. Petersson and J. Tabaczek,
Nucl. Phys. B {\bf 495}, 463 (1997).
\bibitem{grav2} Bas V. de Bakker, Phys. Lett. B {\bf 389}, 238 (1996).
\bibitem{wealth} Z. Burda, D. Johnston, J. Jurkiewicz, M. Kami\'nski, 
M. A. Nowak, G. Papp and I. Zahed, Phys. Rev. E {\bf 65}, 026102 (2002).
\bibitem{cn1} P. L. Krapivsky, S. Redner and F. Leyvraz, 
Phys. Rev. Lett. {\bf 85}, 4629 (2000).
\bibitem{bbw} P. Bialas, Z. Burda and B. Waclaw, 
AIP Conf. Proc. {\bf 776}, 14 (2005). 
\bibitem{haged1} R. Hagedorn, Nuovo Cim. Suppl. 
{\bf 3}, 147 (1965); Nuovo Cim. A {\bf 56}, 1027 (1968).
\bibitem{haged2} Ph. Blanchard, S. Fortunato and 
H. Satz, Eur. Phys. J. C {\bf 34}, 361 (2004).
\bibitem{shaken} J. Eggers, Phys. Rev. Lett. {\bf 83}, 5322 (1999).
\bibitem{shaken2} A. Lipowski and M. Droz, Phys. Rev. E {\bf 65} 
031307 (2002).
\bibitem{kac} T. H. Berlin and M. Kac, Phys. Rev. {\bf 86}, 821 (1952).
\bibitem{god} C. Godr\'{e}che and J. M. Luck, J. Phys. A {\bf 38},
7215 (2005).
\bibitem{cc4} S. Grosskinsky and T. Hanney, Phys. Rev. E {\bf 72},
016129 (2005). 
\bibitem{jdn} J. D. Noh, Phys. Rev. E {\bf 72}, 056123 (2005). 
\bibitem{ournew} B. Waclaw, L. Bogacz, Z. Burda and W. Janke, 
in preparation.

\end{thebibliography}
\end{document}